\documentclass[twocolumn,showpacs,preprintnumbers,amsmath,amssymb,prl,superscriptaddress,aps,10pt]{revtex4-1}

\usepackage{graphicx}
\usepackage{float}
\usepackage{overpic}
\usepackage[usenames,dvipsnames]{color}

\begin{document}

\title{Detecting an exciton crystal by statistical means}
\author{D. Breyel}
\affiliation{Institut f\"ur Theoretische Physik,
Ruprecht-Karls-Universit\"at Heidelberg,\\
 Philosophenweg 19, D-69120 Heidelberg, Germany}
\author{H. Soller}
\affiliation{Institut f\"ur Theoretische Physik,
Ruprecht-Karls-Universit\"at Heidelberg,\\
 Philosophenweg 19, D-69120 Heidelberg, Germany}
\author{T. L. Schmidt}
\affiliation{Departement Physik, Universit\"at Basel, Klingelbergstrasse 82, 5056 Basel, Switzerland}
\author{A. Komnik}
\affiliation{Institut f\"ur Theoretische Physik,
Ruprecht-Karls-Universit\"at Heidelberg,\\
 Philosophenweg 19, D-69120 Heidelberg, Germany}
\date{\today}

\begin{abstract}
We investigate an ensemble of excitons in a coupled quantum well excited via an applied laser field. Using an effective disordered quantum Ising model, we perform a
numerical simulation of the experimental procedure and calculate the probability distribution function $P(M)$ to create $M$ excitons as
well as their correlation function. It shows clear evidence of the existence of two phases corresponding to a liquid and a crystal phase. We demonstrate that not 
only the correlation function but also the distribution $P(M)$ is very well suited to monitor this transition.
\end{abstract}


\maketitle

The exciton is a very fascinating composite particle, which can be generated and investigated in specially designed semiconductor heterostructures -- the bilayer systems. In its simplest incarnation it is a bound state of an electron and a hole and is thus of bosonic nature.
If the system size of such a compound is small it is natural to ask, whether a Bose-Einstein condensation (BEC) of such structures is possible  \cite{keldysh1,*PhysRev.126.1691}. On the other hand, also the possibility of a Cooper-pair-like ground state [the conventional Bardeen-Cooper-Schrieffer (BCS) superconductivity] has been considered in a number of works \cite{comte,*keldysh2,*lozovik}.
Recent experimental progress in the field of electronic bilayer systems allowed for a study
of all these different fascinating possibilities \cite{eisenstein1,*snoke,*su2008,*PhysRevLett.104.027004,*PhysRevLett.68.1383,*PhysRevLett.108.156401,*PhysRevLett.90.226804,gossard,*2012arXiv1210.3176A}.

In most experimental realizations the holes and electrons are spatially separated so that every such indirect exciton has a relatively large dipole moment. It turns out that at some intermediate density, at which the BEC condensation is prohibited, the dipole interactions let the excitons see each other. If the correlations are strong enough even a long-ranged ordering of Wigner crystal type is possible  \cite{springerlink:10.1134/1.1826178,*PhysRevB.78.045313,*PhysRevB.75.155314,0295-5075-72-3-396,*PhysRevB.80.195313,*lozovik2}. A detection of such kind of crystalline structure is very difficult though. The excitons themselves are usually generated with the help of laser fields and they are rather fragile with respect to irradiation. Thus traditional spectroscopic techniques are very difficult to apply and one needs alternative methods \cite{springerlink:10.1134/1.1826178}.  One such approach is based on the knowledge of the first order correlation function, the measurement of which was very recently reported in Ref.~\cite{gossard,*2012arXiv1210.3176A}.

Here we propose an alternative statistical method of detecting and analyzing the properties of the exciton crystallization phenomenon and discuss its predictive power.
A typical experimental cycle would start with the generation of excitons via a laser pulse in a coupled quantum well structure (e.g. GaAs/AlGaAs heterostructure).
After that the number of excitons $M$ is measured by their recombination. Conducting a large number of cycles one gathers the statistics of $M$
\cite{PhysRevB.76.085304,PhysRevLett.73.304}. The fact that for a given $M$ a regular arrangement of the excitons on a lattice minimizes their interaction
energy should be visible in the probability distribution $P(M)$. Although  the correlation function possesses a higher predictive power, we shall show below
that the crystallization can even be seen in $P(M)$, which is accessible by much less effort.

\begin{figure}[ht]
\includegraphics[width=0.7\columnwidth]{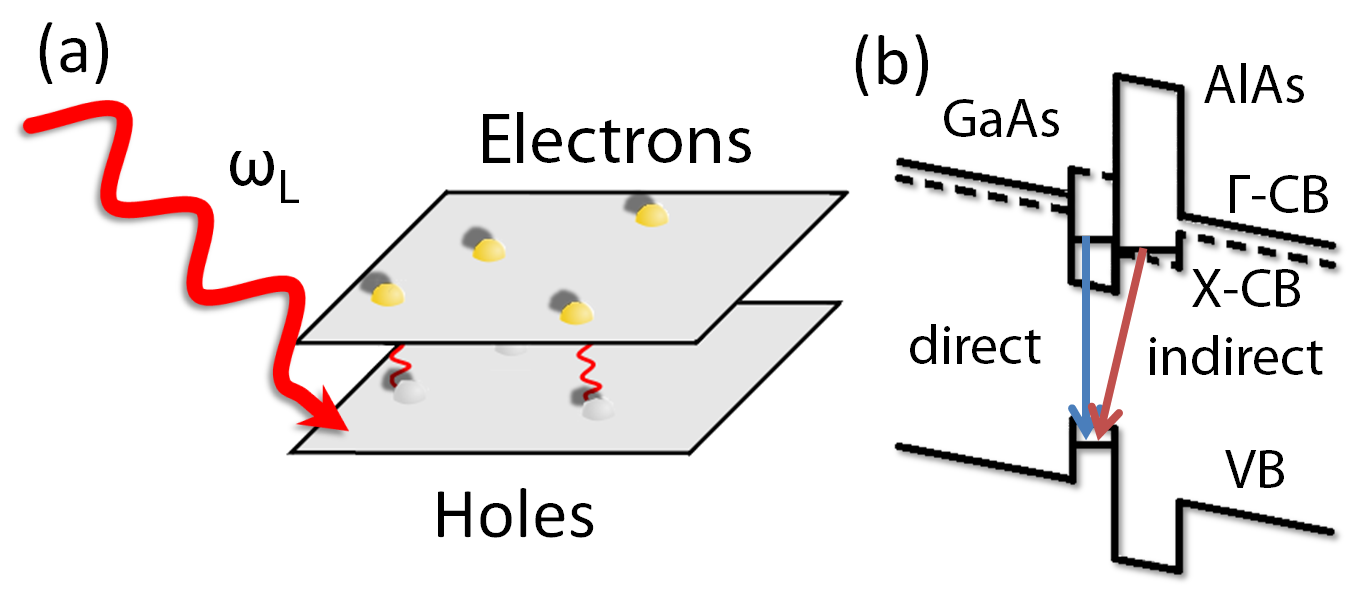}
\caption{(a) shows the typical experimental situation: excitons are created in a coupled quantum well via an applied laser field that excites electrons from the
valence band into the conduction band and thus forms excitons. Possible exciton states in a typical coupled quantum well such as GaAs/AlAs are shown in (b).
These coupled quantum wells have two conduction bands ($\Gamma$ and $X$) originating from different points in the Brillouin zone and allow for both the
formation of direct and indirect excitons, however, with different binding energies.}
\label{fig1}
\end{figure}

We assume that before every measurement cycle the system consists of $N$ valence band electrons, located at random positions $\mathbf{r}_i$. We model each electron
by a two-level system consisting of an unexcited state and an exciton state. A laser field with frequency $\omega_L$ excites the system. It is detuned from
the actual transition frequency between these states by $\Delta$. By means of the laser intensity the Rabi frequency $\omega$, which accounts for
transitions from ground to exciton state, can be changed.
In a typical experiment \cite{2002Natur.418..754S} one uses short laser pulses with high intensity, for which the Rabi frequency is of
the same order as the exciton binding energy \cite{PhysRevB.48.17811}. A circularly polarized light beam (polarization $\sigma$) at a suitably chosen
frequency can create direct and indirect neutral excitons with well defined spins $\sigma$ and $\bar{\sigma} = - \sigma$ by exciting an electron from
the valence band of the same or the other semiconductor in the coupled quantum well (layer), respectively  (see Fig. \ref{fig1}) \cite{PhysRevLett.106.107402,PhysRevLett.73.304}.

In the following we will focus only on the indirect excitons, the particle and hole of which are located in different layers. The formation of direct excitons can be
neglected, since either their formation can be suppressed by an appropriate choice of the excitation frequency  $\omega_L$, or one can simply wait long enough -- their recombination time is much shorter. The indirect excitons possess dipole moments $d$ perpendicular to the layers $d = e D$, where $D$ is the interlayer separation, see Fig. \ref{fig1}(a).

We treat the exciton as a two level system \cite{kowalik-seidl:011104} with electrons either in the valence band or excited to an exciton
\cite{PhysRevLett.98.060405}. We assume the levels to be sharp neglecting effects from the Fermi distribution of the separate bands. We can do so in the
limit of a large detuning of the laser from the resonance \cite{PhysRevB.48.17811}. The electrons interact with the laser light and with each other
through the dipole interaction when they form excitons. The velocity distribution of excitons in coupled quantum wells \cite{2002Natur.418..754S}
can be tuned by efficient cooling \cite{gossard,*2012arXiv1210.3176A} and the application of a perpendicular magnetic field, which gives rise to a higher effective mass
\cite{PhysRevLett.73.304,PhysRevB.65.235304,*springerlink:10.1007/BF00114330}. With both the laser pulse duration \cite{PhysRevLett.94.226401} and
the experimental measurement taking $\tau < 0.1$ ns and the resulting velocities $v_{Ex} \approx 10$ m/s the displacement of a single exciton
is $v_{Ex} \tau < 1$ nm. The typical separations of the excitons one usually encounters are of the order $100$ nm so that we can assume the excitons to
be fixed in space \cite{Exciton_size}. The size of an exciton is estimated via its Bohr radius $a_B \approx 20$nm, see e.g. \cite{kuznetsova:201106}. Since the electrostatic properties do not depend on the details of exciton states we model the system as a randomly arranged
interacting ensemble of spin 1/2 sub-systems each representing a single electron/exciton. Hence, the Hamiltonian reads \cite{PhysRevA.72.063403,PhysRevLett.98.060405}
\begin{eqnarray}
H &=& - \frac{\Delta}{2} \sum_{i=1}^N \sigma_z^{(i)} + \frac{\omega^{*}}{2} \sum_{i=1}^N \sigma_x^{(i)} \nonumber\\  \label{Model_1}
&& + \frac{C^{*}}{4} \sum_{i=2}^{N} \sum_{j=1}^{i-1} \frac{(1+ \sigma_z^{(i)})(1+ \sigma_z^{(j)})}{|\mathbf{r}_i - \mathbf{r}_j|^3}, \label{hexciton}
\end{eqnarray}
where $\sigma_{x,z}^{(i)}$ denote the Pauli matrices. This equation uses the rotating wave approximation when describing the light-matter interaction as in
Ref.~\cite{PhysRevB.38.3342}. It neglects all terms oscillating with frequencies $\omega_L$ and higher.
From here on we measure all energies in units of $\Delta$ entailing $\omega = \omega^{*}/\Delta$. The interaction strength is therefore measured in units of $\Delta L^{3}$, leading to the
definition $C = C^{*} / \Delta L^{3}$. The ground
state particle density $n=N/V$, where $V$ is either a one-dimensional interval or a two-dimensional square, is given for all plots.

The model (\ref{Model_1}) can be interpreted as a (generalised) spin-1/2  Ising model. The Rabi frequency $\omega^{*}$ and the detuning $\Delta$ correspond to magnetic
fields in $x$- and $z$-direction. The third parameter $C^{*}$ indicates the strength of the effective interaction between the excitons. We should note that a similar model
has originally been applied to interactions between Rydberg atoms, where a similar crystal-like phase exists
\cite{PhysRevB.65.235304,natigor,*2012arXiv1203.2884G,*2012arXiv1203.4341B}. In the case of Rydberg atoms, induced dipole moments gave rise to
van-der-Waals interactions. In the case of excitons, in contrast, dipole-dipole interactions between the excitons lead to a stronger dependence
of the distance, $\propto |r|^{-3}$.

We are only interested in the case $\Delta > 0$, because otherwise it is energetically not favorable to produce excitons.
We typically use a large detuning from the transition frequency in accordance with experimental studies \cite{2002Natur.418..754S}, so that $\Delta$ is larger but still
comparable in magnitude to the Rabi frequency. A typical laser field has an excitation frequency of  about 1 eV. The excitons have a dipole moment oriented perpendicular to the plane. In this case $C^{*} = e^2 D^2 / \epsilon$. For the dielectric spacer between the top and bottom layer we assume $\epsilon = 12.9 \epsilon_0$ being a typical
value for $\mbox{GaAs}$ and $D=11.5$nm as put forward by \cite{PhysRevB.85.165452}. 
In the numerical simulations we take the length $L$ of the simulated square
in 2D to be $\approx 200 - 500$nm. For such and larger system dimensions we did not detect any sizeable finite size effects.

The numerical procedure emulates the experimental process by initially generating a random distribution of $N$ electrons [spins in Eq.~(\ref{Model_1})] in a given
one- or two-dimensional volume with open or periodic boundary conditions.
Then  the corresponding Hamiltonian matrix is set up. For large $N$ the size of this matrix is reduced by the truncation of the Hilbert space, which is done by
taking into account  only  $k$ basis states with the smallest diagonal elements in the Hamiltonian matrix. We systematically checked that all the results
do not depend on $k$.
In the next step the eigenvector corresponding to the smallest eigenvalue, the ground state (GS), is calculated by matrix diagonalization \footnote{As was shown
in experiments on Rydberg systems the true ground state is routinely achieved using, for example, chirped laser pulses, see Ref.~\cite{0953-4075-44-18-184008}. }.
It is given as a
linear combination of the previously mentioned basis states and therefore enables an efficient evaluation of different observables. Besides the number of
excitons (which can be non-integer since the ground state is generally a superposition) the pair correlation function $g(r)$, which is closely related to
the density distribution of excitons, can be computed in the following way: we divide the possible range for distances between electrons in our system into
equidistant bins and measure the distance between each pair of electrons to assign it to a certain bin. For each pair the squares of coefficients from the representation of the GS as a linear combination are summed over those states in which the particular pair of electrons is excited. The sum over
multiple random arrangements then produces the correlation function.
\begin{figure}[ht]
\begin{center}
\begin{overpic}[width=0.8\columnwidth]{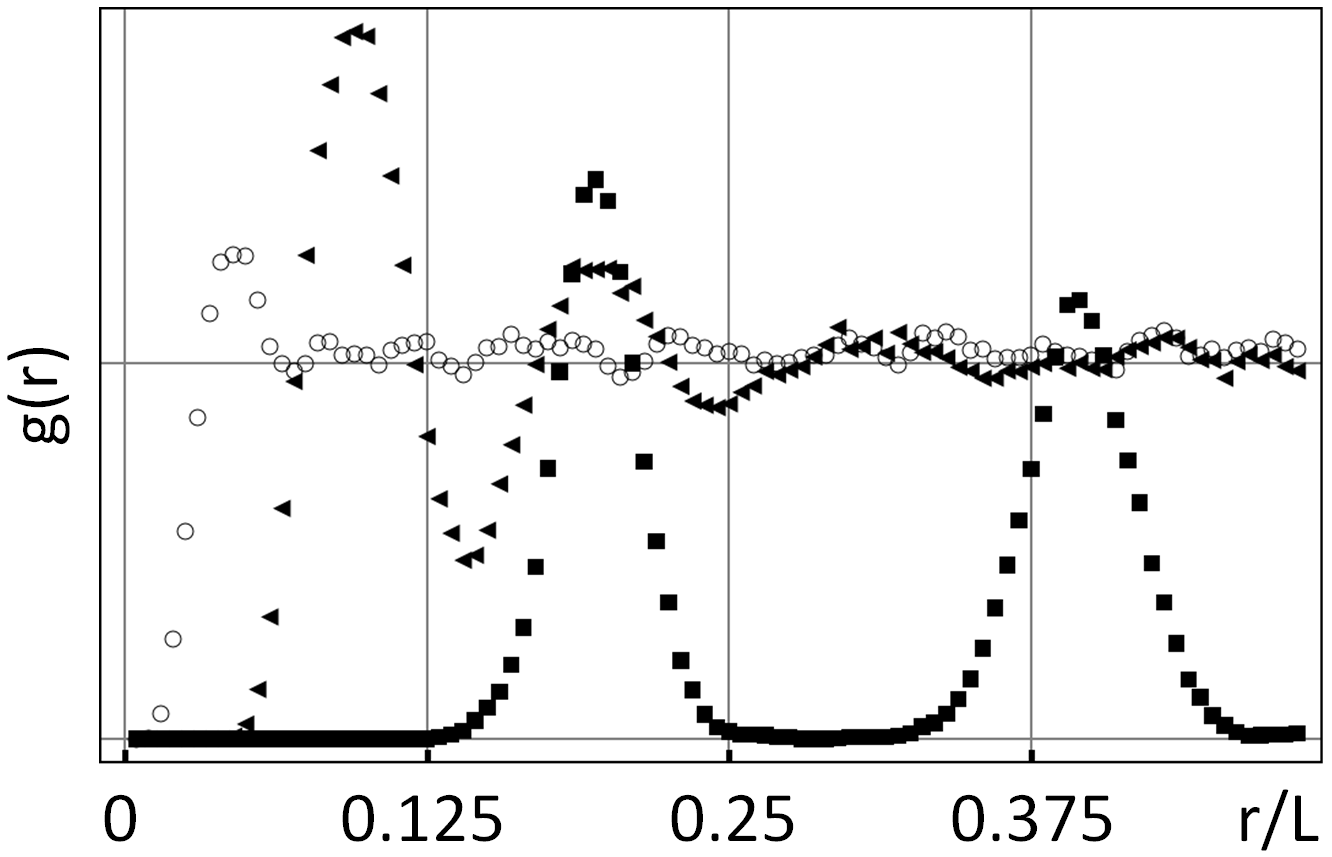}
\end{overpic}
 \caption{Correlation function for the 1D case for $\omega = 0.1$ and $nL=25$ for different interactions strengths. The open circles correspond to $C=7 \cdot 10^{-5}$,
	  the black triangles refer to $C=7 \cdot 10^{-4}$ and the black squares are for $C=7 \cdot 10^{-3}$. 
	  }
 \label{fig:correlationex2}
\end{center}
\end{figure}
\begin{figure}[ht]
\includegraphics[width=0.8\columnwidth]{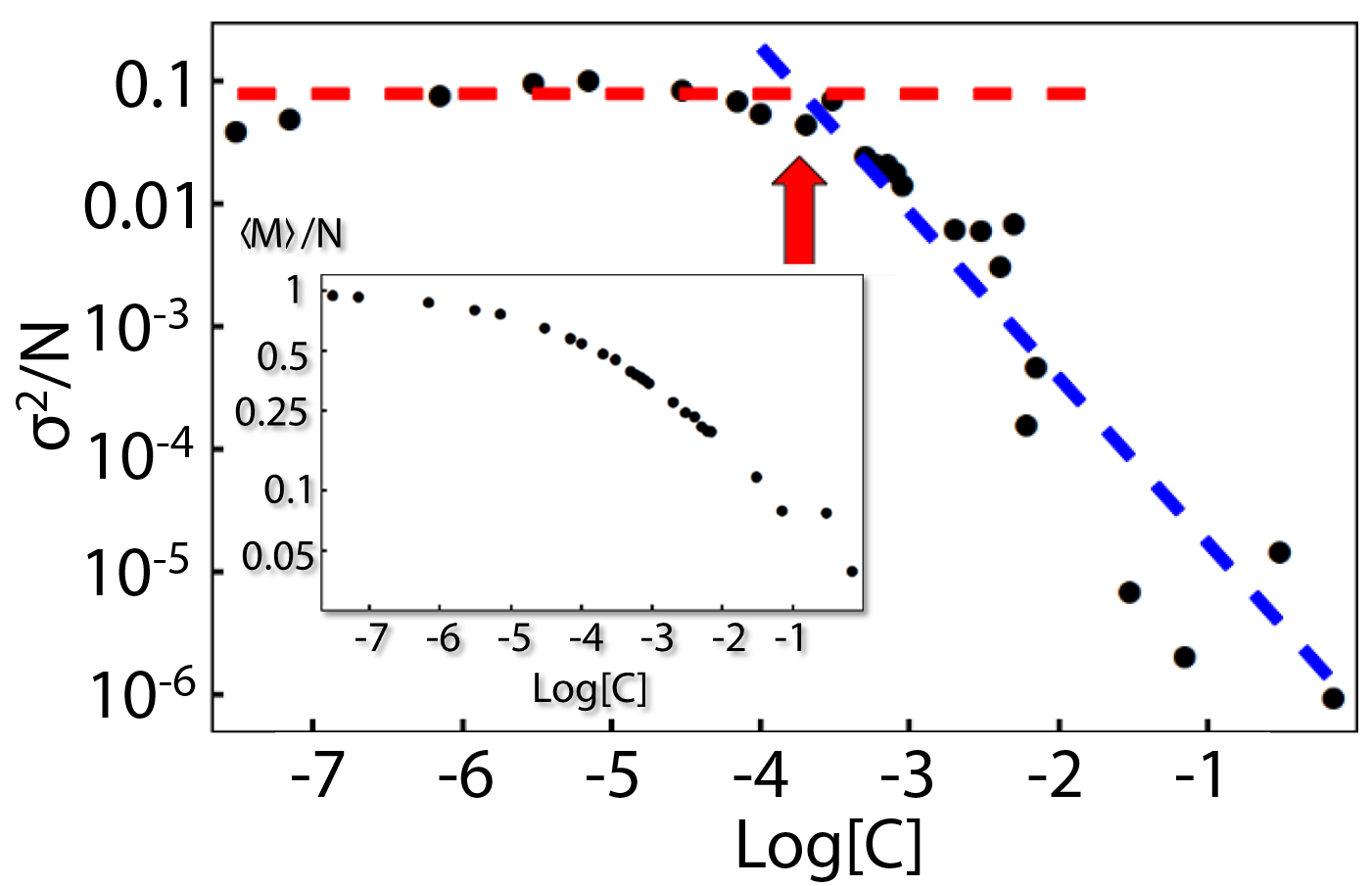}
 \caption{Variance of the mean number of excitons for $\omega=0.1$, and $nL=25$. The straight lines and the arrow illustrate the phase transition. The inset shows the excited fraction of excitons, which is bounded by 1 in the limit of weak interaction.}
\label{fig:mean}
\end{figure}

We start by considering a 1D system with different interaction strengths. Already for weak interactions we see that the correlation
function is zero for $r<R_B$, where $R_B$ can be interpreted as the blockade radius, see Fig.~ \ref{fig:correlationex2}. This effect is due to strong dipole field in vicinity of an exciton which suppresses
the excitation
of additional ones.  As expected $R_B$ grows for increasing interaction strength. Simultaneously an emergence of peaks of higher order is observed. For instance,
at $C ~\gtrsim  10^{-4}$ the second order maximum is already very pronounced and the pair correlation function resembles that of a liquid. Finally,
at $C > C_{\rm crit} \approx 10^{-3}$ $g(r)$ drops to almost zero between the peaks, which we attribute to the crystalline ordering
 of the excitons which is
caused by `bubbles' around each exciton in which no further excitation is possible. 

It is instructive to analyze how this evolution of the system is reflected in the distribution $P(M)$. The simplest distribution parameters are the mean $\langle M \rangle$ and the variance $\sigma^2 = \langle M^2 \rangle - \langle M \rangle^2$, which we plot in Fig.~\ref{fig:mean}. Especially the latter quantity significantly changes its behaviour in vicinity of the critical interaction strength. For weaker $C$ there is only very little change in $\sigma^2$ while for $C>C_{\rm crit}$ it is described by a
power law. This drastic change is due to formation of the dipole crystalline order. Thus, just by measuring
$P(M)$ one can identify the interaction strength at which the crystallization takes place.

The system under consideration features long-ranged interactions so that the coordination number for every particle is effectively infinite. That is why it is reasonable
to expect that the actual dimensionality of the system should not play any significant role in the physical picture. In order to substantiate this argument we have
extended our analysis to 2D systems. Here one has to work with a larger number of particles. As a result one is confronted with considerably longer computation times (they grow exponentially with the number of particles). Nonetheless one is able to perform the same numerical program as in a 1D case.

\begin{figure}[b]
\begin{center}
\includegraphics[width=0.8\columnwidth]{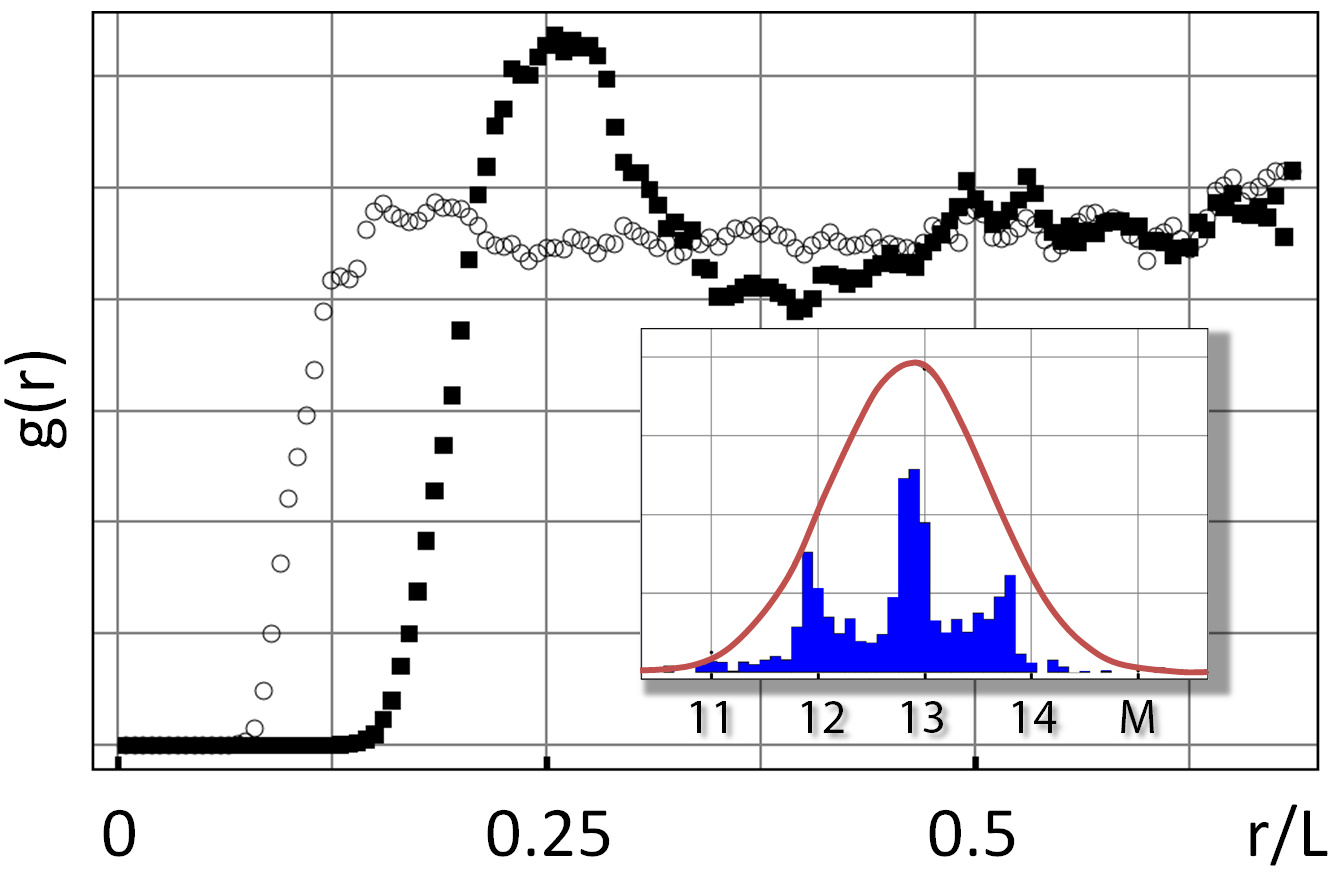}
\caption{Pair correlation function for $\omega=0.1$, $C = 8 \cdot 10^{-3}$ at $nL^2=25$ (black squares). The blockade radius as well as the first two maxima of the curve are clearly visible. The double peak of the second maximum really is a single peak and only splits due to commensurability with the periodic boundary conditions in the sample shown. The data points for the largest distances are less reliable than the rest since only very few runs contribute to these points. We compare
our result to the case $C=2 \cdot 10^{-3}$ where no additional peaks and only the blockade radius is present (open circles). The inset shows the histogram of the number
of excitons with a Gaussian fit for the same parameters as for $g(r)$.}
\label{fig:correlation}
\end{center}
\end{figure}

\begin{figure*}
\includegraphics[width=0.8\textwidth]{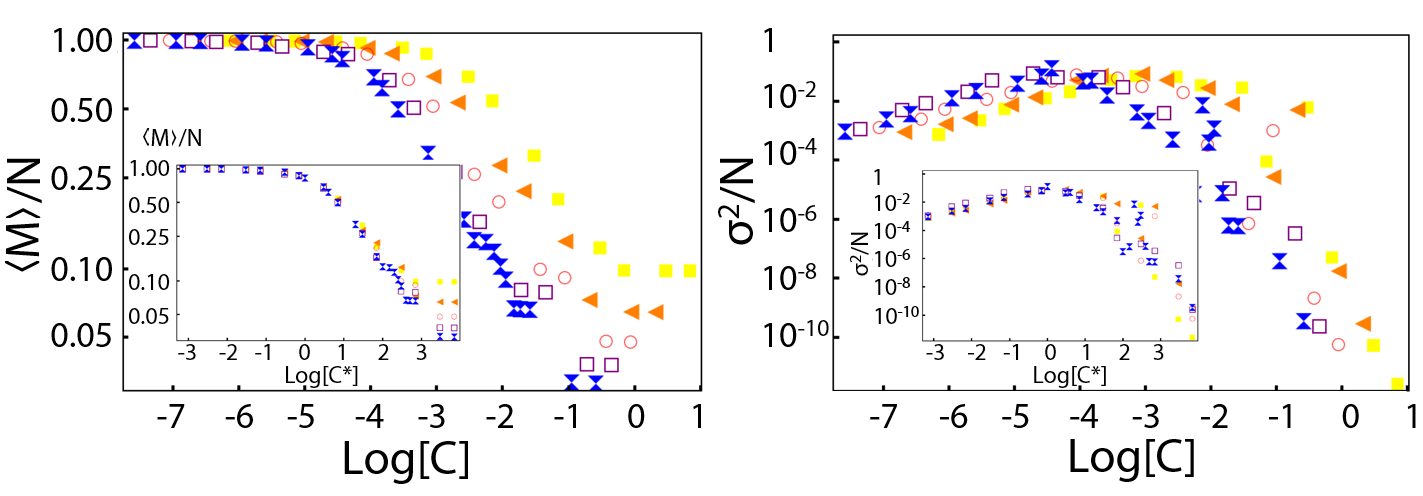}
\caption{Excited fraction (left panel) and variance (right panel) of a 2D system (this is a direct analogon of Fig.~3 of the 1D case). Different data sets correspond 
	  to different particle numbers at equal density. The coding is as follows: $N=10$ (yellow, solid squares), $N=15$ (orange, solid triangles), $N=20$ (red, open 
	  circles), $N=25$ (purple, open squares) and $N=30$ (blue, sand clocks). One observes a very good convergence for increasing $N$. Taking the edge length of the
	  system with $N=30$ as a reference one can map all the curves on it by rescaling of $L$, which is done in the insets by streching the abscissa by a factor 
	  of $(L_{N=30}/L_N)^3$.}
 \label{fig:2Dmeanandvariance}
\end{figure*}

In Fig.~\ref{fig:correlation} we plot the correlation function in 2D case, which is generated for a system with periodic boundary condition and 30 spins.
For weak interactions we again obtain a `flat' correlation function, which is essentially uniform for $r>R_B$. For $C>C_{\rm crit} \approx 8 \cdot 10^{-3}$ $g(r)$ starts
to develop additional maxima, which is a signature of the crystallization onset.
As in the 1D case we would like to simultaneously monitor the evolution of $P(M)$, see Fig.~\ref{fig:2Dmeanandvariance} for the plot of the average and variance.
The variance $\sigma^2$ [as well as $g(r)$] shows a clear transition at $C \approx 10^{-3}$. The change of the power law decay of $M$ is not
as abrupt as in another possible realisation of a Wigner crystal though \cite{2012arXiv1203.4341B}.
The remarkable change of the variance as a function of $C$ shows that also simple statistical analysis of the exciton number allows to identify the phase
transition to an exciton crystal.

In conclusion we have investigated the formation and detection of an exciton crystal in a typical semiconductor environment using exact numerical diagonalization and
approximative descendants of this method for one- and two-dimensional systems. Both the pair correlation function and the simpler statistical data show signatures of
this phase transition. The blockade phenomenon distinguishes the crystal- and liquid-like behavior from the BEC phase, whereas additional peaks in $g(r)$ show the 
phase transition from the liquid to the crystal phase.

The authors would like to thank F. Dolcini and S. Maier for many interesting discussions. AK is supported by CQD and 'Enable fund' of the University of Heidelberg.

\bibliography{exciton}
\end{document}